\documentclass[12pt]{article}
\usepackage{amsmath}
\usepackage{amssymb}
\usepackage{graphicx}
\usepackage{float}
\usepackage{indentfirst}
\usepackage{mathrsfs}
\usepackage{dsfont}

\usepackage{setspace}

\usepackage[labelfont=bf,labelsep=period]{caption}

\usepackage[numbers,sort&compress]{natbib}

\newtheorem{theorem}{Theorem}

\newtheorem{corollary}{Corollary}

\newtheorem{observation}{Observation}
\title{Separable Decompositions of Bipartite Mixed States}

\author
{Jun-Li Li and Cong-Feng Qiao$^{\ast}$\\ [0.2cm]
\normalsize{Department of Physics, University of the Chinese Academy of Sciences,}\\
\normalsize{YuQuan Road 19A, Beijing 100049, China}\\[2pt]
\normalsize{Key Laboratory of Vacuum Physics, University of Chinese Academy of Sciences}\\[3mm]
\normalsize{$^\ast$ To whom correspondence should be addressed; E-mail: qiaocf@ucas.ac.cn.}
}

\date{}
\begin{document}
\baselineskip24pt \maketitle
\vspace{0.5cm}
\begin{abstract} \doublespacing
We present a practical scheme for the decomposition of a bipartite mixed state into a sum of direct products of local density matrices, using the technique developed in Li and Qiao (Sci. Rep. 8: 1442, 2018). In the scheme, the correlation matrix which characterizes the bipartite entanglement is first decomposed into two matrices composed of the Bloch vectors of local states. Then we show that the symmetries of Bloch vectors are consistent with that of the correlation matrix, and the magnitudes of the local Bloch vectors are lower bounded by the correlation matrix. Concrete examples for the separable decompositions of bipartite mixed states are presented for illustration.
\end{abstract}

\newpage

\section{Introduction}

Entanglement lies at the heart of quantum information theory. The qualitative and quantitative studies of entanglement are not only of great importance to our understanding of quantum theory, but also have practical applications in quantum computation and quantum information processing \cite{NC-Book}. A prior question in the study of quantum entanglement is to determine whether a given quantum state is entangled or not. A mixed bipartite state of particles $A$ and $B$ is separable if and only if it can be expressed as \cite{Mixed-Ent-Criterion}
\begin{equation}
\rho_{AB} = \sum_{i=1}^L p_i \rho_i^{(A)} \otimes \rho_i^{(B)} \; . \label{Gen-Sep-exp}
\end{equation}
Here $p_i>0$ with $\sum_{i=1}^L p_i = 1$, and $\rho_i^{(A)}$ and $\rho_i^{(B)}$ are local density matrices of the particles $A$ and $B$. Unlike the pure state, the separability of a mixed state is computationally hard to be determined, even for the bipartite system \cite{NP-separable}.

One remarkable criterion in detecting the entanglement is the positive partial transposition (PPT) criterion \cite{PPT-criterion}; however it is necessary and sufficient only for systems of $2\times 2$ and $2\times 3$ \cite{22and23}. Many practical criteria have been developed ever since, whereas being either necessary or sufficient, unfortunately. These criteria may be roughly sorted into two classes. One involves inequalities of computational norms which may be regarded as scalar measures of the entanglement \cite{cnc, realignement, Bloch-Rep, Recent-norms}. Violations of these inequalities indicate the existence of entanglement. Another class is based on the expectation values of some appropriately chosen observables, named entanglement witness \cite{22and23,LURs, Sperling-Vogel, Recent-LURs}. By exploring a complete set of observables, there were also the attempts to unify these two classes \cite{covariance-matrix, Recent-cova}. In a recent work \cite{Separability-QL}, we introduced the multiplicative Horn's inequalities to the separability problem of bipartite states. Though in principle the criterion in \cite {Separability-QL} is necessary and sufficient, its physical significance and practical applications need to be exemplified.

In this work, we develop a series of practical methods for the decomposition of a bipartite state into the sum of direct products of two local states based on the technique in \cite{Separability-QL}. With the help of Bloch representation of the quantum state, the correlation matrix of bipartite state is first decomposed into product of two factor matrices. Then, by considering the magnitudes and symmetries of the singular values and singular vectors of the factor matrices, practical entanglement criteria can be obtained. Remarkably, the separable decompositions of bipartite mixed states can be constructed explicitly based on our criteria. In the end of this work, neat examples are presented as applications of the method.

\section{The separable decompositions of bipartite states}

\subsection{The Bloch representation of a quantum state}

An arbitrary $N\times M$ dimensional bipartite state in the Bloch representation is
\begin{eqnarray}
\rho_{AB} & = & \frac{1}{NM}  \mathds{1} \otimes \mathds{1} + \frac{1}{2M} \vec{a} \cdot \vec{\lambda}\otimes \mathds{1} + \frac{1}{2N} \mathds{1} \otimes \vec{b} \cdot \vec{\sigma} + \frac{1}{4} \sum_{\mu=1}^{N^2-1} \sum_{\nu=1}^{M^2-1} \mathcal{T}_{\mu\nu} \, \lambda_{\mu} \otimes \sigma_{\nu} \; , \label{rhoAB-Bloch-Gen}
\end{eqnarray}
where $\mathds{1}$ is the identity matrix, $\vec{a}$ and $\vec{b}$ have the components of $a_{\mu} = \mathrm{Tr}[\rho_{AB} (\lambda_{\mu}\otimes \mathds{1})]$ and $b_{\nu} = \mathrm{Tr}[ \rho_{AB} (\mathds{1} \otimes \sigma_{\nu})]$, and the correlation matrix $\mathcal{T}_{\mu\nu} = \mathrm{Tr}[\rho_{AB}(\lambda_{\mu} \otimes \sigma_{\nu})]$. The vector $\vec{\lambda}$ in equation (\ref{rhoAB-Bloch-Gen}) is defined to be $\vec{\lambda} \equiv (\lambda_1, \cdots, \lambda_{N^2-1})^{\mathrm{T}}$ with $\lambda_{\mu}$ being the generators of SU($N$), and $\vec{\sigma}$ is defined similarly with $\sigma_{\nu}$ being the generators of SU$(M)$. For example, the three generators of SU($2$) are Pauli matrices
\begin{equation}
\lambda_1 =
\begin{pmatrix}
0 & 1 \\
1 & 0
\end{pmatrix} \; , \; \lambda_2 =
\begin{pmatrix}
0 & -i \\
i & 0
\end{pmatrix}\; , \; \lambda_3 =
\begin{pmatrix}
1 & 0 \\
0 & -1
\end{pmatrix} \; ,
\end{equation}
while the generators are the eight Gell-Mann matrices for $N=3$. The reduced density matrices for particles $A$ and $B$ are obtained from $\rho_{AB}$ via the following
\begin{equation}
\rho_A = \mathrm{Tr}_B[\rho_{AB}] = \frac{1}{N} \mathds{1} + \frac{1}{2}\vec{a} \cdot \vec{\lambda} \; , \; \rho_B = \mathrm{Tr}_A[\rho_{AB}] = \frac{1}{M} \mathds{1} + \frac{1}{2}\vec{b} \cdot \vec{\sigma} \; .
\end{equation}
Here $\vec{a}$ and $\vec{b}$ are called the Bloch vectors of the density matrices. A bipartite state is separable if it can be decomposed as the sum of direct products of local density matrices as shown in equation (\ref{Gen-Sep-exp}). The necessary and sufficient condition for the separability of $\rho_{AB}$ in equation (\ref{rhoAB-Bloch-Gen}) reads \cite{Separability-QL}
\begin{eqnarray}
\sum_{i=1}^L p_i \vec{r}_i = \vec{a} \; , \; \sum_{j=1}^L p_j \vec{s}_j = \vec{b} \; , \;
\sum_{k=1}^{L} p_k \vec{r}_k \vec{s}_k^{\,\mathrm{T}} = \mathcal{T}  \; , \label{General-form-density}
\end{eqnarray}
where $p_i>0$, $\sum_{i=1}^L p_i=1$, and $\rho_i^{(A)} = \frac{1}{N} \mathds{1} + \frac{1}{2} \vec{r}_i \cdot \vec{\lambda}$ and $\rho_i^{(B)} = \frac{1}{M} \mathds{1} + \frac{1}{2} \vec{s}_i \cdot \vec{\sigma}$ with $\vec{r}_i$, $\vec{s}_j$ bing the Bloch vectors of the decomposed local quantum states. $L$ in equation (\ref{General-form-density}) stands for the number of local states needed in the separable decomposition. Equation (\ref{General-form-density}) may be expressed in the matrix form
\begin{equation}
M_r\vec{p} = \vec{a} \; , \; M_s \vec{p} = \vec{b} \; , \; M_{rp} M_{sp}^{\mathrm{T}} = \mathcal{T} \; . \label{Mat-Dec-Mrs}
\end{equation}
Here $M_{r} = (\vec{r}_1,\vec{r}_2,\cdots, \vec{r}_L)$ and $M_s = (\vec{s}_1,\vec{s}_2,\cdots, \vec{s}_L)$ with $\vec{r}_i$ and $\vec{s}_j$ being $N^2-1$ and $M^2-1$ dimensional real vectors respectively; $\vec{p} = (p_1,\cdots, p_{L})^{\mathrm{T}}$ and $M_{rp} = M_r D_{p}^{\frac{1}{2}}$, $M_{sp} = M_s D_{p}^{\frac{1}{2}}$ with $D_p = \mathrm{diag}\{p_1,p_2,\cdots, p_L\}$.

As being Hermitian, the reduced density matrices can be unitarily diagonalized as
\begin{align}
\rho'_A & = U_A \rho_{A} U_A^{\dag} = \mathrm{diag}\{\lambda_1^{(A)},\cdots, \lambda_n^{(A)}, 0, \cdots,0\}\; , \label{diag-rhoa} \\ \;
\rho'_B & = U_B \rho_{B} U_B^{\dag} = \mathrm{diag}\{\lambda_1^{(B)},\cdots,\lambda_m^{(B)}, 0, \cdots,0\} \; , \label{diag-rhob}
\end{align}
where $\lambda_i^{(A)}$ and $\lambda_i^{(B)}$ are positive real numbers, and $n$ and $m$ represent the ranks of the reduced density matrices (local ranks). Because the state $\rho'_{AB} = (U_A\otimes U_B) \rho_{AB} (U_A^{\dag} \otimes U_B^{\dag})$ has the same separability as $\rho_{AB}$, we have the following observation
\begin{observation}
For the state $\rho_{AB}$ whose local ranks are $n$ and $m$, if the correlation matrix $\mathcal{T}'_{\mu\nu}$ of $\rho'_{AB}$ has nonzero elements for $\mu>n^2-1$ or $\nu>m^2-1$, then $\rho_{AB}$ is entangled. \label{Observation-1}
\end{observation}

\noindent{\bf Proof:} Suppose $\rho'_{AB}$ is separable, then $\rho'_{AB} = \sum_{i} p_i\rho_{i}^{(A)} \otimes \rho_{i}^{(B)}$, and
\begin{equation}
\rho'_A = \sum_{i}p_i \rho_i^{(A)} \; , \;  \rho'_B = \sum_{i}p_i \rho_i^{(B)} \; . \label{rank-normal}
\end{equation}
Here $p_i>0$ with $\sum_i p_i=1$; $\rho_A'$, $\rho_i^{(A)}$, $\rho_B'$, and $\rho_{i}^{(B)}$ are all positive semidefinite matrices. According to equation (\ref{diag-rhoa}), $\rho_i^{(A)}$ in equation (\ref{rank-normal}) can only take the following form:
\begin{equation}
\rho_i^{(A)} =
\begin{pmatrix}
X_{n\times n} & 0 \\
0 & 0
\end{pmatrix}_{N\times N}\; .
\end{equation}
This is because the diagonal elements of positive semidefinte matrices must be nonnegative, so we have $(\rho_{i}^{(A)})_{kk} = 0$ for $k >n$. Furthermore, from the row and column inclusion properties we have: if $(\rho_{i}^{(A)})_{kk} = 0$, then $(\rho_{i}^{(A)})_{\mu k} = (\rho_{i}^{(A)})_{k\mu} = 0$ for all $\mu \in \{1,\cdots, N\}$ (Observation 7.1.10 of \cite{Matrix-analysis}). Hence, the Bloch vectors $\vec{r}_i$ of $\rho_i^{(A)}$ are
\begin{equation}
\rho_i^{(A)} = \left(\frac{1}{n} \mathds{1} + \frac{1}{2}\sum_{\mu =1}^{n^2-1} r_{i\mu} \lambda_{\mu} \right)_{n\times n} \oplus \mathbf{0}_{(N-n) \times (N-n)} \; ,
\end{equation}
where $r_{i\mu}$ are components of $\vec{r}_i$ and lie in the Bloch vector space of SU($n$) $\subset$ SU($N$). Similar arguments apply to $\rho_i^{(B)}$ as well. That means, if $\rho'_{AB}$ is separable, then $\mathcal{T}'_{\mu\nu}=0$ for $\mu>n^2-1$ or $\nu>m^2-1$. This completes the proof. Q.E.D.

A straightforward corollary of Observation \ref{Observation-1} goes as follows:
\begin{corollary}
All $N\times M$ mixed states with local ranks $n<N$ and $m<M$ are either reducible to $n\times m$ bipartite states with full local ranks, or entangled.
\end{corollary}
Therefore we need only to consider the separability problem for mixed bipartite states whose reduced density matrices have full local ranks. The full local rank state could be further transformed into a normal form with maximally mixed subsystems, where the normal form is separable or entangled only when the original state is separable or entangled \cite{normal-form}. The normal form of a bipartite state $\rho_{AB}$ is expressed as
\begin{equation}
\rho_{AB} \mapsto \widetilde{\rho}_{AB} = \frac{1}{NM} \mathds{1} \otimes \mathds{1} + \frac{1}{4} \sum_{\mu=1}^{N^2-1}\sum_{\nu=1}^{M^2-1} \widetilde{T}_{\mu\nu} \lambda_{\mu} \otimes \sigma_{\nu} \; .
\end{equation}
Hereafter in this paper, the bipartite state $\rho_{AB}$ is assumed to be in its normal form, i.e. the Bloch representation of $\rho_{AB}$ has $\vec{a}=0$ and $\vec{b}=0$.

\subsubsection*{Local symmetries: linear maps on Bloch vectors}

The vectorization of a matrix $A\in \mathbb{C}^{N\times M}$ is defined as:
\begin{equation}
\mathcal{V}(A) \equiv (A_{11},\cdots,A_{N1},A_{12},\cdots, A_{N2},\cdots,A_{1M},\cdots, A_{NM})^{\mathrm{T}} \; .
\end{equation}
The following transformation induces a linear map on $\rho$, $\hat{S}: \rho \mapsto \hat{S}(\rho)$,
\begin{equation}
\hat{S}(\rho) \equiv \mathcal{W}[X \mathcal{V}(\rho)] \;, \label{realization-linear}
\end{equation}
where $X \in \mathbb{C}^{NM\times NM}$ is an $NM\times NM$ matrix with complex elements, and $\mathcal{W} \equiv \mathcal{V}^{-1}$ is the inverse operation of vectorization which wraps a vector into a matrix \cite{Four-partite-SLOCC}. It is easy to verify that $\hat{S}$ is linear, i.e. $\hat{S}(a\rho_1+b\rho_2) = a \hat{S}(\rho_1) + b\hat{S}(\rho_2)$. With properly chosen $X$, we can get
\begin{equation}
\rho = \frac{1}{N}\mathds{1} + \frac{1}{2} \vec{r} \cdot \vec{\lambda} \mapsto \hat{S}(\rho) = \frac{1}{N} \mathds{1} + \frac{1}{2} \vec{r}\,' \cdot \vec{\lambda} \; . \label{S-realize}
\end{equation}
Here $\vec{r}\,' = O\vec{r}$ and the matrix $O \in \mathbb{R}^{(N^2-1) \times (N^2-1)}$ is induced by $X$. Note, while $\hat{S}(\rho)$ is Hermitian and trace one, it may not be positive semidefinite.

We define the matrix realignment operation to an $I_1\cdot I_2\times I_1\cdot I_2$ dimensional matrix $A$ as \cite{Realignement}
\begin{equation}
\mathcal{R}(A) \equiv (\mathcal{V}(A_{11}), \cdots, \mathcal{V}(A_{I_11}), \mathcal{V}(A_{12}), \cdots, \mathcal{V}(A_{I_12}), \cdots, \mathcal{V}(A_{I_1I_1})) \; ,
\end{equation}
where $A_{ij}$ are $I_2\times I_2$ submatrices of $A$. Linear operations acting on the local states of a bipartite density matrix may be realized via the following transformation:
\begin{equation}
\hat{S}_{A} \otimes \hat{S}_B(\rho_{AB}) \equiv \mathcal{R}^{-1}[\mathcal{W}( X_A \otimes X_B \cdot \mathcal{V}[\mathcal{R} (\rho_{AB})] )] \; . \label{line-op-define}
\end{equation}
Here $\hat{S}_A(\rho_A) = \mathcal{W}[X_A \mathcal{V}(\rho_A)]$, $\hat{S}_B(\rho_B) = \mathcal{W}[X_B \mathcal{V}(\rho_B)]$, $\mathcal{R}^{-1}$ is the inverse of $\mathcal{R}$, and
\begin{equation}
\hat{S}_A(\rho_A) = \frac{1}{N} \mathds{1} + (O_A\vec{r}\, ) \cdot \vec{\lambda} \; , \; \hat{S}_B(\rho) = \frac{1}{M} \mathds{1} + (O_B\vec{s}\, ) \cdot \vec{\sigma} \; ,
\end{equation}
with $O_{A,B}$ being induced by $X_{A,B}$, respectively. The linear operation $\hat{S}_{A} \otimes \hat{S}_B$ maps the normal form $\rho_{AB}$ to the following
\begin{align}
\hat{S}_{A} \otimes \hat{S}_B(\rho_{AB}) = \frac{1}{NM}  \mathds{1} \otimes \mathds{1} + \frac{1}{4} \sum_{\mu,\nu}\mathcal{T}'_{\mu\nu} \, \lambda_{\mu} \otimes \sigma_{\nu}\; ,
\end{align}
where $\mathcal{T}' = O_A \mathcal{T} O_B^{\mathrm{T}}$, and $O_A \in \mathbb{R}^{(N^2-1) \times (N^2-1)}$, $O_B \in \mathbb{R}^{(M^2-1) \times (M^2-1)}$.

\subsubsection*{Magnitudes: singular values of the factor matrices}

Applying the singular value decomposition, we have
\begin{equation}
M_{rp} = R^{(1)} \Lambda_{\alpha} Q^{(1)}  \; , \; M_{sp} = R^{(2)} \Lambda_{\beta} Q^{(2)} \; . \label{M_rsq-singular}
\end{equation}
Here $R^{(1)} \in \mathrm{SO}(N^2-1)$, $R^{(2)} \in \mathrm{SO}(M^2-1)$, and $Q^{(1)}, Q^{(2)} \in \mathrm{SO}(L)$. Taking $M_{rp}$ as example, the singular value matrix $\Lambda_{\alpha} \in \mathbb{R}^{(N^2-1)\times L}$ has the following form
\begin{align}
\Lambda_{\alpha} & =
\begin{pmatrix} D_{\alpha} \\ 0 \end{pmatrix} \in \mathbb{R}^{(N^2-1)\times L} \; , \mathrm{if}\ N^2-1>L \; ,  \\
D_{\alpha} & =
\begin{pmatrix} \Lambda_{\alpha} \\ 0 \end{pmatrix} \in \mathbb{R}^{L\times L} \; , \mathrm{if}\ N^2-1<L \; ,
\end{align}
where $D_{\alpha} = \mathrm{diag}\{\alpha_1, \cdots, \alpha_L\}$ and $D_{\alpha} = \Lambda_{\alpha}$ for $L=N^2-1$. Similar formulation applies to $\Lambda_{\beta}$ and $D_{\beta}$ as well. Let $\alpha_1\geq \alpha_2 \geq \cdots \geq \alpha_n >0$ be the $n$ nonzero singular values of $M_{rp}$, $\beta_1\geq \beta_2 \geq \cdots \geq \beta_m >0$ be the $m$ nonzero singular values of $M_{sp}$, and $\tau_1\geq \cdots \geq \tau_l >0$ be the $l$ nonzero singular values of $\mathcal{T}$, then from the decomposition $\mathcal{T} = M_{rp}M_{sp}^{\mathrm{T}}$ we have the Sylvester's rank inequality: $(n+m-L)\leq l \leq \mathrm{min}\{n,m\} \leq \mathrm{max}\{n,m\} \leq L$.

The necessary and sufficient criterion presented in \cite{Separability-QL} may be summarized as two steps: 1. The existence of the decomposition of equation (\ref{Mat-Dec-Mrs}); 2. The decomposition can be realized in the physical region of Bloch vectors (that is, $\rho_i^{(A)}$ and $\rho_i^{(B)}$ in equation (\ref{General-form-density}) must be positive semidefinite). Here we provide practical procedures to exemplify these steps: $M_{rp}$ and $M_{sp}$ must have appropriate singular values and singular vectors.

\subsection{Correlation matrix decomposition}

Let $\mathcal{T} = (\vec{u}_1,\cdots,\vec{u}_{N^2-1}) \Lambda_{\tau} (\vec{v}_1, \cdots, \vec{v}_{M^2-1})^{\mathrm{T}}$ be the singular value decomposition of the correlation matrix $\mathcal{T}$ and $\Lambda_{\tau}$ has rank $l$, then we have
\begin{equation}
\mathcal{T} = \sum_{\mu =1}^l \tau_{\mu} \vec{u}_{\mu}\vec{v}_{\mu}^{\,\mathrm{T}} \; . \label{Singular-tau-def}
\end{equation}
For the $l$ nonzero values of $\tau_{\mu}$, the corresponding singular vectors $\{\vec{u}_1,\cdots,\vec{u}_l\}$ and $\{\vec{v}_1,\cdots,\vec{v}_l\}$ span two $l$-dimensional subspaces in Bloch vector space: $\mathcal{S}_l^{(A)} \equiv \mathrm{span}\{\vec{u}_1,\cdots,\vec{u}_l\} \subseteq \mathcal{S}_{N^2-1}$ and $\mathcal{S}_l^{(B)} \equiv \mathrm{span} \{\vec{v}_1,\cdots,\vec{v}_l\} \subseteq \mathcal{S}_{M^2-1}$. Let $D_{\tau} = \mathrm{diag}\{\tau_1,\cdots,\tau_l,0,\cdots, 0\}$ be an $L\times L$ diagonal matrix, then $M_{rp}$ and $M_{sp}$ in equation (\ref{Mat-Dec-Mrs}) can always be expressed as \cite{Separability-QL}
\begin{align}
M_{rp} & =  M_{r}D_{p}^{\frac{1}{2}} = (\vec{u}_1,\cdots, \vec{u}_L) X D_{\alpha} Q^{(1)} \; , \label{Theorem-eq-Mr}\\
M_{sp} & = M_{s}D_{p}^{\frac{1}{2}} = (\vec{v}_1,\cdots, \vec{v}_L) Y D_{\beta} Q^{(2)} \; ,  \label{Theorem-eq-Ms}
\end{align}
where $X$, $Y$, $Q^{(1,2)}$ are orthogonal matrices, $\vec{u}_{\mu}$ and $\vec{v}_{\nu}$ are the left and right singular vectors of $\mathcal{T}$, and $D_{\tau}$ has the same singular values as $D_{\alpha}  Q^{(1)} Q^{(2)\mathrm{T}} D_{\beta}^{\mathrm{T}}$ according to Theorem 1 of \cite{Separability-QL}. Note that the value of $L$ may be larger than $N^2-1$ (and/or $M^2-1$), in which case we shall regard $M_{r} = (\vec{r}_1, \cdots, \vec{r}_L)$ as having $L$-dimensional column vectors of $\vec{r}_i = (r_{i1},\cdots, r_{iN^2-1}, 0, \cdots,0)^{\mathrm{T}}$. The Frobenius norm of a matrix $M$ is $||M||_{2} \equiv ( \mathrm{Tr} [MM^{\mathrm{T}}])^{\frac{1}{2}}$, and we have the following theorem
\begin{theorem}
If the correlation matrix of a bipartite state can be decomposed as $\mathcal{T} = M_{rp}M_{sp}^{\mathrm{T}}$, then
\begin{equation}
||\mathcal{T}||_{2}^2 = \vec{\boldsymbol{\alpha}}^{\,\mathrm{T}} \mathcal{Q}\, \vec{\boldsymbol{\beta}}\; . \label{double-stocha}
\end{equation}
Here $\vec{\boldsymbol{\alpha}} = (\alpha_1^2,\cdots, \alpha_L^2)^{\mathrm{T}}$, $\vec{\boldsymbol{\beta}} = (\beta_1^2,\cdots, \beta_L^2)^{\mathrm{T}}$, and $\mathcal{Q}$ is an $L\times L$ orthostochastic matrix; $\alpha_i$ and $\beta_j$ are singular values of $M_{rp}$ and $M_{sp}$ in descending order. \label{Theorem-1}
\end{theorem}

\noindent{\bf Proof:} Equations (\ref{Theorem-eq-Mr}) and (\ref{Theorem-eq-Ms}) lead to $M_{rp}^{\mathrm{T}} M_{rp} = D^{\frac{1}{2}}_{p} M_{r}^{\mathrm{T}} M_{r} D^{\frac{1}{2}}_{p} = Q^{(1)\mathrm{T}} D_{\alpha}^2 Q^{(1)}$ and $M_{sp}^{\mathrm{T}} M_{sp} = D^{\frac{1}{2}}_{p} M_s^{\mathrm{T}} M_s D^{\frac{1}{2}}_{p} = Q^{(2)\mathrm{T}} D_{\beta}^2 Q^{(2)}$. It can be shown that ( 5.03a of Ref. \cite{Matrix-topics})
\begin{equation}
\begin{pmatrix}
|\vec{r}_1|^2p_1 \\
  \vdots \\
|\vec{r}_L|^2p_L
\end{pmatrix} = \mathcal{Q}_1 \vec{\boldsymbol{\alpha}} \; , \;
\begin{pmatrix}
|\vec{s}_1|^2p_1 \\
  \vdots \\
|\vec{s}_L|^2p_L
\end{pmatrix} = \mathcal{Q}_2 \vec{\boldsymbol{\beta}} \; . \label{Cor-maj-eq}
\end{equation}
Here $\mathcal{Q}_1 = Q^{(1)\mathrm{T}}\circ Q^{(1)\mathrm{T}}$, $\mathcal{Q}_2 = Q^{(2)\mathrm{T}} \circ Q^{(2)\mathrm{T}}$ are orthostochastic matrices  \cite{Book-Majorization} with $(A\circ B)_{ij} \equiv A_{ij}B_{ij}$ been the Hadamard product of two matrices. Let $Q = Q^{(1)}Q^{(2)\mathrm{T}}$,
we have
\begin{equation}
||\mathcal{T}||_{2}^2 = \mathrm{Tr}[M_{rp}M_{sp}^{\mathrm{T}}M_{sp}M_{rp}^{\mathrm{T}}] = \mathrm{Tr}[D_{\alpha}^2 Q D_{\beta}^2 Q^{\mathrm{T}}] = \vec{\boldsymbol{\alpha}}^{\mathrm{T}} \mathcal{Q}\, \vec{\boldsymbol{\beta}} \; , \label{L1-eq-tr}
\end{equation}
where Lemma 5.1.5 of \cite{Matrix-topics} is used in the last equality and $\mathcal{Q}$ is an orthostochastic matrix  with $\mathcal{Q}_{ij} = Q_{ij}^2$. (An orthostochastic matrix is also a doubly stochastic matrix).  Q.E.D.

In equation (\ref{Cor-maj-eq}), the sums of the components of the left hand sides of the equalities may be considered as the mean squared norms of the Bloch vectors of the decomposed local states
\begin{equation}
\mathcal{E}(A) \equiv \sum_{i=1}^L p_i |\vec{r}_i|^2 \; , \; \mathcal{E}(B) \equiv \sum_{i=1}^L p_i |\vec{s}_i|^2 \; . \label{Quan-Mea}
\end{equation}
Because the norm of a Bloch vector is related to the purity of the quantum state, i.e., $|\vec{r}\,|^2 = 2(\mathrm{Tr}[\rho^2] - \frac{1}{N})$, the quantities $\mathcal{E}(A,B)$ in equation (\ref{Quan-Mea}) may be regarded as the mean ``quantumness" of the decomposed states. Because the mean of Bloch vectors is $ \sum_{i=1}^L p_i\vec{r}_i = \vec{a}$, we are legitimate to define the variance of the Bloch vectors $\Delta (A)^2 \equiv \mathcal{E}(A) - |\vec{a}\,|^2$ where $\vec{a} = 0$ for normal form bipartite states. In this sense, the following quantities
\begin{equation}
\Delta(A)^2 = \mathcal{E}(A) \; , \; \Delta(B)^2 = \mathcal{E}(B)
\end{equation}
may be regarded as the fluctuations of the local Bloch vectors' distributions.

We define the average of the squares of the components along the directions $\vec{u}_{\mu} \in \mathcal{S}^{(A)}_{l}$ and $\vec{v}_{\nu} \in \mathcal{S}^{(B)}_{l}$ as follows
\begin{equation}
\mathcal{E}_{\mu}(A) \equiv \sum_{i=1}^L p_i |\vec{u}_{\mu} \cdot \vec{r}_i|^2 \; ,\; \mathcal{E}_{\nu}(B) \equiv \sum_{i=1}^L p_i |\vec{v}_{\nu} \cdot \vec{s}_i|^2 \; .
\end{equation}
The Ky Fan norm of a matrix $\mathcal{T}$ is defined to be the sum of its singular values, i.e. $||\mathcal{T}||_{\mathrm{KF}} \equiv \sum_{\mu=1}\tau_{\mu}$, and we have the following as our main separability criterion
\begin{corollary}
The squared norms of the local states' Bloch vectors are lower bounded by the correlation matrix in the following way
\begin{align}
\left(\sum_{\mu=1}^l \mathcal{E}_{\mu} (A) \right) \cdot \left(\sum_{\nu=1}^l \mathcal{E}_{\nu} (B) \right) & \geq ||\mathcal{T}||_{\mathrm{KF}}^2 \; , \label{cor-sum-un} \\
\left( \prod_{\mu=1}^l \mathcal{E}_{\mu} (A) \right) \cdot \left( \prod_{\nu=1}^l \mathcal{E}_{\nu}(B) \right) & \geq \prod_{\mu=1}^l \tau_{\mu}^2 \; . \label{cor-prod-un}
\end{align}
Here $\mathcal{E}_{\mu} (A)$ and $\mathcal{E}_{\nu} (B)$ are the means of the squares of the components along the unit directions $\vec{u}_{\mu} \in \mathcal{S}_l^{(A)}$ and $\vec{v}_{\nu} \in \mathcal{S}_l^{(B)}$, respectively, and $\tau_{\mu}$ are the singular values of $\mathcal{T}$. \label{Cor-uncertainty}
\end{corollary}

\noindent{\bf Proof:} According to the decomposition of equations (\ref{Theorem-eq-Mr}, \ref{Theorem-eq-Ms}) we have
\begin{align}
\mathcal{T} = M_{rp} \cdot M_{sp}^{\mathrm{T}} = (\vec{u}_1,\cdots, \vec{u}_L) X D_{\alpha} Q^{(1)} \cdot Q^{(2)\mathrm{T}} D_{\beta}^{\mathrm{T}} Y^{\mathrm{T}} \begin{pmatrix}
\vec{v}^{\,\mathrm{T}}_1 \\
\vdots \\
\vec{v}^{\,\mathrm{T}}_L
\end{pmatrix}\; ,\label{Ms-Gram}
\end{align}
where we made an explicit separation between the matrices by a dot product. There exists a real orthogonal matrix $Q_1 \in$ SO($L$) such that
\begin{equation}
\mathcal{T} =  M_{rp} Q_1  Q_1^{\mathrm{T}} M_{sp}^{\mathrm{T}} = M_{rp} Q_1
\begin{pmatrix}
\mathds{1}_{l\times l} & 0 \\
0 & 0
 \end{pmatrix} Q_1^{\mathrm{T}} M_{sp}^{\mathrm{T}} \; . \label{full-rank-form}
\end{equation}
This can be shown by the following. Choosing $Q_1 = Q^{(1)\mathrm{T}}
\begin{pmatrix}Q''_{n\times n} & 0 \\ 0 & \mathds{1}_{(L-n)\times (L-n)}\end{pmatrix}$ we have
\begin{align}
M_{rp}Q_1 = (\vec{u}_1,\cdots, \vec{u}_L) X
\begin{pmatrix}
D_{n\times n} Q''_{n\times n} & 0 \\
0 & 0
\end{pmatrix} \; ,
\end{align}
where $Q_{n\times n}'' \in$ SO($n$), and $D_{n\times n} = \mathrm{diag}\{\alpha_1, \cdots, \alpha_n\}$ has $\alpha_1\geq \cdots \geq \alpha_n>0$.  According to the full-rank factorization of a matrix (section 0.4.6 (e,f) of Ref. \cite{Matrix-analysis}) the first $n$ rows of $Q_1^{\mathrm{T}}M_{sp}^{\mathrm{T}}$ must have the same rank $l$ as $\mathcal{T}$. Because $l\leq n$, an appropriate choice of $Q''_{n\times n}$ would satisfy equation (\ref{full-rank-form}).

The singular value decomposition of the first $l$ columns of $M_{rp}Q_1$ and $M_{sp}Q_1$ reads
\begin{align}
M_{rp}Q_1 & = U'(\Lambda_{\alpha'},\vec{r}\,'_{\!\! l+1},\cdots, \vec{r}\,'_{\!\!n}, 0_{n+1},\cdots, 0_{L})
\begin{pmatrix}
\overline{Q}^{(1)}_{l\times l} & 0 \\
0 & \mathds{1}_{(L-l)\times (L-l)}
\end{pmatrix} \; , \\
M_{sp}Q_1 & = V'(\Lambda_{\beta'},0_{l+1}, \cdots, 0_{n}, \vec{s}\,'_{\!\! n+1},\cdots, \vec{s}\,'_{\!\!L})
\begin{pmatrix}
\overline{Q}^{(2)}_{l\times l} & 0 \\
0 & \mathds{1}_{(L-l)\times (L-l)}
\end{pmatrix} \; .
\end{align}
Here $\Lambda_{\alpha'}$ has the form of $\begin{pmatrix} D_{\alpha'} \\ 0\end{pmatrix}$ and so is $\Lambda_{\beta'}$. $D_{\tau'} = \mathrm{diag} \{\tau_1,\cdots,\tau_l\}$ must be the singular value matrix for $D_{\alpha'} \overline{Q}^{(1)}_{l\times l} \overline{Q}^{(2)\mathrm{T}}_{l\times l} D_{\beta'}$, i.e., $D_{\tau'} = \overline{X} D_{\alpha'} \overline{Q}^{(1)}_{l\times l} \overline{Q}^{(2)\mathrm{T}}_{l\times l} D_{\beta'} \overline{Y}^{\mathrm{T}}$, where $\overline{X}, \overline{Y} \in$ SO($l$), $D_{\alpha'} =\mathrm{diag}\{\alpha_1',\cdots,\alpha_l'\}$, and $D_{\beta'} =\mathrm{diag}\{\beta_1',\cdots,\beta_l'\}$. The left and right singular vectors, $U' = \{\vec{u}\,'_{\!\!1}, \cdots, \vec{u}\,'_{\!\!l},\cdots\}$ and $V' = \{\vec{v}\,'_{\!\! 1},\cdots, \vec{v}\,'_{\!\!l},\cdots\}$, have the following relations with that of $\mathcal{T}$ in equation (\ref{Singular-tau-def})
\begin{align}
\{\vec{u}\,'_{\!\!1},\cdots, \vec{u}\,'_{\!\!l}\} & = \{\vec{u}_{1},\cdots, \vec{u}_{l}\}\overline{X} \; , \\
\{\vec{v}\,'_{\!\!1},\cdots, \vec{v}\,'_{\!\!l}\} & = \{\vec{v}_{1},\cdots, \vec{v}_{l}\} \overline{Y} \; ,
\end{align}
Define the projection $\vec{u}_{\mu} \vec{u}_{\mu}^{\,\mathrm{T}}$, $\mu \in \{1,\cdots, l\}$, then
\begin{align}
\mathcal{E}_{\mu} (A) & = \mathrm{Tr}[ M_{rp}^{\mathrm{T}} \vec{u}_{\mu} \vec{u}_{\mu}^{\mathrm{T}} M_{rp} ] = \mathrm{Tr}[ Q_1^{\mathrm{T}}M_{rp}^{\mathrm{T}} \vec{u}_{\mu} \vec{u}_{\mu}^{\mathrm{T}} M_{rp} Q_1 ] \nonumber \\
& = \sum_{\nu =1}^l \overline{X}_{\mu\nu}^2\alpha'^2_{\nu} + c_{\mu}^2 \geq \sum_{\nu=1}^l \overline{X}_{\mu\nu}^2 \alpha'^2_{\nu} \; . \label{epsilon-mu-eq}
\end{align}
Here $c_{\mu}^2$ represents the sum of squared components of $\overline{X} \{\vec{r}\,'_{\!\! l+1}, \cdots, \vec{r}\,'_{\!\! n}\}$ along $\vec{u}_{\mu}$. Define $\varepsilon_{\mu}^2 \equiv \sum_{\nu =1}^l \overline{X}_{\mu\nu}^2 \alpha'^2_{\nu}$, we have the following relation
\begin{align}
\begin{pmatrix}
\varepsilon_1^2 \\
\varepsilon_2^2 \\
\vdots \\
\varepsilon_l^2
\end{pmatrix} =
\begin{pmatrix}
\overline{X}_{11}^2 & \overline{X}_{12}^2 & \cdots & \overline{X}_{1l}^2 \\
\overline{X}_{21}^2 & \overline{X}_{22}^2 & \cdots & \overline{X}_{2l}^2 \\
\vdots & \vdots & \ddots & \vdots \\
\overline{X}_{l1}^2 & \overline{X}_{l2}^2 & \cdots & \overline{X}_{ll}^2
\end{pmatrix}
\begin{pmatrix}
\alpha'^2_1 \\
\alpha'^2_2 \\
\vdots \\
\alpha'^2_l
\end{pmatrix} \; ,
\end{align}
which may be expressed as $\vec{\boldsymbol{\varepsilon}} = \overline{\mathcal{X}} \, \vec{\boldsymbol{\alpha}'}$. Here $\vec{\boldsymbol{\varepsilon}} = (\varepsilon_1^2,\cdots,\varepsilon_l^2)^{\mathrm{T}}$, $\vec{\boldsymbol{\alpha}}' = (\alpha'^2_1,\cdots,\alpha'^2_l)^{\mathrm{T}}$, and  $\overline{\mathcal{X}}$ is a doubly stochastic matrix. Considering equation (\ref{epsilon-mu-eq}), we have the following
\begin{align}
\sum_{\mu=1}^l \mathcal{E}_{\mu} (A) & \geq \sum_{\mu=1}^l \varepsilon_{\mu}^2 = \sum_{\mu=1}^l \alpha'^2_{\mu} \; ,  \label{sum-lower-bound}\\
\prod_{\mu=1}^l \mathcal{E}_{\mu} (A) & \geq \prod_{\mu=1}^l \varepsilon_{\mu}^2 \geq \prod_{\mu=1}^l \alpha'^2_{\mu} \; , \label{prod-lower-bound}
\end{align}
where the last equality and inequality in equations (\ref{sum-lower-bound}) and (\ref{prod-lower-bound}) are properties of doubly stochastic matrix, and similar relations exist for $\mathcal{E}_{\nu}(B)$. Taking the Schwartz inequality $(\sum_i \alpha'^2_i) (\sum_{j}\beta'^2_j) \geq (\sum_k \alpha'_k\beta'_k)^2$ and the multiplicative Horn's inequalities \cite{Separability-QL}, equations (\ref{sum-lower-bound}, \ref{prod-lower-bound}) lead to
\begin{align}
\left(\sum_{\mu=1}^l \mathcal{E}_{\mu} (A) \right) \left(\sum_{\nu=1}^l \mathcal{E}_{\nu}(B) \right) & \geq \left( \sum_{\mu=1}^l \tau_{\mu} \right)^2 \; , \\
\left(\prod_{\mu=1}^l \mathcal{E}_{\mu}(A) \right) \left( \prod_{\nu=1}^l \mathcal{E}_{\nu} (B)\right) & \geq \left(\prod_{\mu=1}^l \alpha'^2_{\mu} \right) \left(\prod_{\nu=1}^l \beta'^2_{\nu} \right) = \prod_{\mu =1}^l \tau_{\mu}^2 \; .
\end{align}
Q.E.D.

As the physical region of the Bloch vectors of high dimensional system has symmetric properties \cite{Bloch-N}, there exists the following Corollary
\begin{corollary}
If $\rho_{AB}$ is separable and $\hat{S}_A$ and $\hat{S}_B$ are positive linear maps over the subspaces $\mathcal{S}_l^{(A,B)}$ for all $\rho_A$ and $\rho_B$, then $\hat{S}_A \otimes \hat{S}_B$ are also positive linear maps for $\rho_{AB}$. \label{Cor-sym}
\end{corollary}
\noindent{\bf Proof:} The corollary is quite clear from the following relation
\begin{align}
\hat{S}_A\otimes \hat{S}_B(\rho_{AB}) & = \hat{S}_A \otimes \hat{S}_B \left( \sum_{i}p_i \rho^{(A)}_i \otimes \rho^{(B)}_i \right) \nonumber \\
& = \sum_i p_i \hat{S}_A(\rho^{(A)}_i) \otimes \hat{S}_B(\rho^{(B)}_i) \; .
\end{align}
Here $\hat{S}_{A,B}$ are symmetric operations (discrete or continuous, i.e. reflections, permutations, rotations, scalings, etc.) over the subspaces $\mathcal{S}_l^{(A)}$ and $\mathcal{S}_l^{(B)}$ defined in equation (\ref{line-op-define}). Q.E.D.

For the normal form states with maximally mixed subsystems, the typical known results using the Bloch representation may be expressed as follows \cite{Bloch-Rep}: If the bipartite mixed state is separable then
\begin{equation}
||\mathcal{T}||_{\mathrm{KF}}^2 \leq \frac{4(N-1)(M-1)}{NM}\; , \label{res-other-1}
\end{equation}
and if
\begin{equation}
||\mathcal{T}||_{\mathrm{KF}}^2 \leq \frac{4}{NM(N-1)(M-1)}\; , \label{res-other-2}
\end{equation}
then the state is separable. By considering the squared norms of the local Bloch vectors and their symmetries, our main results lie in Corollaries \ref{Cor-uncertainty} and \ref{Cor-sym}, where the singular vectors and singular values of the correlation matrix $\mathcal{T}$ play more important roles than that of in equations (\ref{res-other-1}) and (\ref{res-other-2}). In the following, we shall show the applications of the entanglement criteria of Corollaries \ref{Cor-uncertainty} and \ref{Cor-sym} and compare them with the existing related ones through neat examples.

\subsection{Examples}

\subsubsection{Example I: The $2\times 4$ PPT entangled state}

Consider the following $2\times 4$ dimensional mixed state
\begin{equation}
\rho_{AB} = \frac{1}{2\cdot 4} \mathds{1}\otimes \mathds{1} + \frac{1}{4} (t_1 \sigma_1 \otimes \lambda_1 + t_2 \sigma_2 \otimes \lambda_{13} + t_3 \sigma_3 \otimes \lambda_{3}) \; , \label{state-24}
\end{equation}
where $t_{\mu} \neq 0$, $t_{\mu} \in \mathbb{R}$, and $\sigma_{\mu}$ and $\lambda_{\nu}$ are SU(2) and SU(4) generators respectively. Equation (\ref{state-24}) represents a physical state when $\rho_{AB}$ is positive semidefinite, that is
\begin{equation}
t_2^2\leq \frac{1}{4} \; , \; (|t_1| + |t_3|)^2 \leq \frac{1}{4} \; . \label{Example-I-p-c}
\end{equation}

{\noindent \bf Existing results:} The density matrix $\rho_{AB}$ in equation (\ref{state-24}) has positive (semidefinite) partial transposition, so the PPT criterion cannot determine whether the state is entangled or separable. The Bloch representation criteria state that: $\rho_{AB}$ is entangled if $||\mathcal{T}||^2_{\mathrm{KF}} > 3/2$, and is separable when  $||\mathcal{T}||^2_{\mathrm{KF}} \leq 1/6$ \cite{Bloch-Rep}. As $||\mathcal{T}||_{\mathrm{KF}} = |t_1|+ |t_2| + |t_3| \leq 1$ for $\rho_{AB}$, the Bloch representation criteria cannot detect the entanglement when
\begin{equation}
\frac{1}{6} < |t_1|+ |t_2| + |t_3| \leq 1 \; .
\end{equation}
And none of the two criteria could be used to construct the separable decompositions for the state $\rho_{AB}$.

\noindent {\bf Our results:} The left and right singular vector spaces of the state $\rho_{AB}$ in equation (\ref{state-24}) are $\mathcal{S}_3^{(A)} = \mathrm{span} \{\sigma_1,\sigma_2,\sigma_3 \}$ and $\mathcal{S}_3^{(B)} = \mathrm{span}  \{\lambda_1, \lambda_{13}, \lambda_3\}$. In order to be positive semidefinite, the Bloch vectors of the one-particle states $\rho^{(A)}$ and $\rho^{(B)}$ cannot have too large components in $\mathcal{S}_3^{(A)}$ and $\mathcal{S}_3^{(B)}$. Evaluations of the positivity conditions of the single particle states in $\mathcal{S}_3^{(A)}$ and $\mathcal{S}_3^{(B)}$ show that
\begin{align}
\sum_{\mu=1}^3 \mathcal{E}_{\mu} (A) \leq 1 \; & , \; \sum_{\nu=1}^3 \mathcal{E}_{\nu} (B) \leq 1 \; , \label{condition-leq}\\
\prod_{\mu=1}^3 \mathcal{E}_{\mu} (A) \leq \frac{1}{27} \; &, \; \prod_{\nu=1}^3 \mathcal{E}_{\nu} (B) \leq \left(\frac{2}{27}\right)^2 \; . \label{condition-geq}
\end{align}
Here, the upper bounds of equation (\ref{condition-leq}) are obtained by the states
\begin{align}
\rho^{(A)}_i & = \frac{1}{2}\mathds{1} + \frac{1}{2}(\sin\theta_i\cos\phi_i \sigma_1 + \sin\theta_i \sin\phi_i \sigma_2 + \cos\theta_i \sigma_3)\; , \\
\rho^{(B)}_i & = \frac{1}{4}\mathds{1} + \frac{1}{2}( \cos\theta_i \lambda_1 + \sin\theta_i \lambda_3 + \frac{1}{\sqrt{3}}\lambda_{8} + \frac{1}{\sqrt{6}}\lambda_{15} )\; .
\end{align}
While the upper bounds of equation (\ref{condition-geq}) are obtained by
\begin{align}
\rho^{(A)} & = \frac{1}{2} \mathds{1} + \frac{1}{2\sqrt{3}}( \pm \sigma_1 \pm \sigma_2 \pm \sigma_3)\; , \\
\rho^{(B)} & = \frac{1}{4}\mathds{1} + \frac{1}{2}(\pm \frac{ \sqrt{2}}{3} \lambda_1 \pm \frac{\sqrt{2}}{3} \lambda_3 + \frac{1}{3}\lambda_{13} + \frac{1}{3\sqrt{3}}\lambda_{8} + \frac{1}{3\sqrt{6}}\lambda_{15} ) \; .
\end{align}
Taking equations (\ref{condition-leq}, \ref{condition-geq}) into Corollary \ref{Cor-uncertainty}, we have that if $\rho_{AB}$ is separable then
\begin{equation}
1\cdot 1 \geq (|t_1|+|t_2| + |t_3|)^2 \; , \; \frac{1}{27} \cdot \frac{4}{27^2} \geq (t_1t_2t_3)^2 \; .
\end{equation}
That is, the state in equation (\ref{state-24}) is entangled when $(|t_1|+|t_2| + |t_3|)^2 > 1$ or $(t_1t_2t_3)^2 > \frac{4}{27^3}$. For the special case of $t_1=t_2=t_3 = t$, we have that $\rho_{AB}$ is entangled when $t>\frac{\sqrt[3]{2}}{3\sqrt{3}} \sim 0.242$, i.e., $t=\frac{1}{4}$ corresponds to an entangled state.

Most importantly, our method can give the separable decomposition of $\rho_{AB}$ within the subspaces $\mathcal{S}_3^{(A)}$ and $\mathcal{S}_3^{(B)}$, i.e. \begin{equation}
\rho_{AB} = \sum_{i=1}^L p_i \rho_{i}^{(A)} \otimes \rho_{i}^{(B)}\; , \label{Example-I-decom}
\end{equation}
where $\rho_i^{(A)} = \frac{1}{2}\mathds{1} + \frac{1}{2}\vec{r}_i\cdot \vec{\sigma}$ and $\rho_{i}^{(B)} = \frac{1}{4}\mathds{1} + \frac{1}{2} \vec{s}_i \cdot \vec{\lambda}$. First, we write the correlation matrix $\mathcal{T}$ of $\rho_{AB}$ as follows (see the proof of Corollary 2 in Ref. \cite{Separability-QL})
\begin{equation}
\mathcal{T} =
\begin{pmatrix}
\alpha_1 & 0 & 0 & 0 \\
0 & \alpha_2 & 0 & 0 \\
0 & 0 & \alpha_3 & 0
\end{pmatrix} Q \cdot Q^{\mathrm{T}} \begin{pmatrix}
\beta_1 & 0 & 0  \\
0 & \beta_2 & 0  \\
0 & 0 & \beta_3 \\
0 & 0 & 0
\end{pmatrix} =M_{rp}\cdot M_{sp}^{\mathrm{T}} \; .
\end{equation}
Here $t_{\mu} = \alpha_{\mu} \beta_{\mu}$, and $Q \in$ SO(4) is chosen to be
\begin{equation}
Q = \frac{1}{2}\begin{pmatrix}
1 & -1 & -1 & 1 \\
-1 & -1 & 1 & 1 \\
-1 & 1 & -1 & 1 \\
1 & 1 & 1 & 1
\end{pmatrix} \; ,
\end{equation}
where we set $\sqrt{p_i} = Q_{4i} = \frac{1}{2}$. Then, decomposed separable states may be expressed as
\begin{align}
M_{rp} & = M_rD_p^{\frac{1}{2}} =
\begin{pmatrix}
  \alpha_1 & -\alpha_1 & -\alpha_1 & \alpha_1 \\
  -\alpha_2 & -\alpha_2 & \alpha_2 & \alpha_2 \\
  -\alpha_3 & \alpha_3 & -\alpha_3 & \alpha_3
\end{pmatrix}
\begin{pmatrix}
\frac{1}{2} & 0 & 0 & 0 \\
0 & \frac{1}{2} & 0 & 0 \\
0 & 0 & \frac{1}{2} & 0 \\
0 & 0 & 0 & \frac{1}{2}
\end{pmatrix} \; , \label{decom-24-rp}\\
 M_{sp} & = M_sD_p^{\frac{1}{2}} =
\begin{pmatrix}
  \beta_1 & -\beta_1 & -\beta_1 & \beta_1 \\
  -\beta_2 & -\beta_2 & \beta_2 & \beta_2 \\
  -\beta_3 & \beta_3 & - \beta_3 & \beta_3
\end{pmatrix}
\begin{pmatrix}
\frac{1}{2} & 0 & 0 & 0 \\
0 & \frac{1}{2} & 0 & 0 \\
0 & 0 & \frac{1}{2} & 0 \\
0 & 0 & 0 & \frac{1}{2}
\end{pmatrix} \; . \label{decom-24-sp}
\end{align}
The corresponding decomposition of $\rho_{AB}$ can be read from equations (\ref{decom-24-rp}) and (\ref{decom-24-sp}), where $p_1=p_2=p_3=p_4=1/4$, and the Bloch vectors for $\rho_i^{(A,B)}$ are
\begin{align}
\vec{r}_1 =
\begin{pmatrix}
\alpha_1 \\
-\alpha_2 \\
-\alpha_3
\end{pmatrix}\; , \; \vec{r}_2 =
\begin{pmatrix}
-\alpha_1 \\
-\alpha_2 \\
\alpha_3
\end{pmatrix}\; , \;
\vec{r}_3 =
\begin{pmatrix}
-\alpha_1 \\
\alpha_2 \\
-\alpha_3
\end{pmatrix}\; , \;
\vec{r}_4 =
\begin{pmatrix}
\alpha_1 \\
\alpha_2 \\
\alpha_3
\end{pmatrix} \; , \label{Example-I-detail-1}\\
\vec{s}_1 =
\begin{pmatrix}
\beta_1 \\
-\beta_2 \\
-\beta_3
\end{pmatrix}\; , \;
\vec{s}_2 =
\begin{pmatrix}
-\beta_1 \\
-\beta_2 \\
\beta_3
\end{pmatrix}\; , \;
\vec{s}_3 =
\begin{pmatrix}
-\beta_1 \\
\beta_2 \\
-\beta_3
\end{pmatrix}\; , \;
\vec{s}_4 =
\begin{pmatrix}
\beta_1 \\
\beta_2 \\
\beta_3
\end{pmatrix} \; . \label{Example-I-detail-2}
\end{align}
In the subspace of $\mathcal{S}_3^{(B)}$, the states $\rho^{(B)} = \frac{1}{4}\mathds{1} + \frac{1}{2}(\beta_1\lambda_1 + \beta_2\lambda_{13} + \beta_3 \lambda_{3})$ require that $\beta_1^2 + \beta_3^2 \leq \frac{1}{4}$ and $\beta_2^2 \leq \frac{1}{4}$. Therefore, $\rho_{AB}$ is separable whenever the following conditions are satisfied
\begin{equation}
\frac{t_1^2}{\alpha_1^2} + \frac{t_3^2}{\alpha_3^2} \leq \frac{1}{4}\; , \; \frac{t_2^2}{\alpha_2^2} \leq \frac{1}{4}\; , \; \mathrm{with}\; \alpha_1^2 + \alpha_2^2+\alpha_3^2 \leq 1 \; ,
\end{equation}
where the region for $t_{\mu}$ forms a cylindroid.

\subsubsection{Example II: The $3\times 3$  octahedral and  tetrahedral states}

We consider the following two $3\times 3$ bipartite states with $t_{\mu} \in \mathbb{R}$ and $t_{\mu} \neq 0$
\begin{align}
\rho_{AB}^{(1)} & = \frac{1}{9} \mathds{1} \otimes \mathds{1} + \frac{1}{4}(t_1\lambda_{1}\otimes \lambda_{1} +t_2 \lambda_{2}\otimes \lambda_{2}+ t_3\lambda_{3}\otimes \lambda_{3} ) \;, \label{rho1-state} \\
\rho_{AB}^{(2)} & = \frac{1}{9} \mathds{1} \otimes \mathds{1} + \frac{1}{4}(t_1\lambda_{1}\otimes \lambda_{1} + t_2\lambda_{2}\otimes \lambda_{4}+ t_3\lambda_{3}\otimes \lambda_{6} )\; .
\end{align}
Here $\rho_{AB}^{(1,2)}$ represent physical states when
\begin{align}
\rho_{AB}^{(1)} \geq 0 & : \hspace{0.5cm} |t_1 \pm t_2|\pm t_3 \leq \frac{4}{9} \; , \label{Example-II-1-p-c}\\
\rho_{AB}^{(2)} \geq 0 & :\hspace{0.5cm} \sqrt{t_1^2 + t_2^2 + t_3^2} \leq \frac{4}{9}\; . \label{Example-II-2-p-c}
\end{align}

\noindent{\bf Existing results:} The PPT criterion detected that $\rho_{AB}^{(1)}$ is entangled when
\begin{equation}
|t_1|+|t_2|+|t_3| > \frac{4}{9} \; . \label{Example-II-1-oct-PPT}
\end{equation}
However the separability within the octahedral region of $|t_1|+|t_2|+|t_3| \leq \frac{4}{9}$ is unknown for PPT criteria. The Bloch representation criteria state that $\rho^{(1)}_{AB}$ is entangled when $||\mathcal{T}||_{\mathrm{KF}} > 4/3$ and is separable when $||\mathcal{T}||_{\mathrm{KF}} \leq 1/3$. A combined application of the PPT and Bloch representation criteria still cannot determine the separability of $\rho_{AB}^{(1)}$ when
\begin{equation}
1/3 < ||\mathcal{T}||_{\mathrm{KF}} = |t_1|+ |t_2| + |t_3| \leq 4/9 \; . \label{Example-II-1-other-res}
\end{equation}
While for the state $\rho_{AB}^{(2)}$, none of the PPT and Bloch representation criteria can determine whether it is entangled or not.

\noindent {\bf Our results:} For state $\rho_{AB}^{(1)}$, the left and right singular vectors of the correlation matrix both are spanned by $\mathcal{S}_{3}^{(A)} =\mathcal{S}_{3}^{(B)} = \mathrm{span} \{\lambda_1,\lambda_2,\lambda_3\}$. There exists the following positive linear map for $\rho = \frac{1}{3}\mathds{1} + \frac{1}{2}(x_1\lambda_1  + x_2 \lambda_2 + x_3 \lambda_3 + \sum_{\mu=4}^8x_{\mu} \lambda_{\mu} )$
\begin{align}
\hat{S} (\rho) = \frac{1}{3}\mathds{1} + \frac{1}{2}(-x_1\lambda_1  - x_2 \lambda_2 - x_3 \lambda_3 + \sum_{\mu=4}^8 x_{\mu} \lambda_{\mu} )\; ,
\end{align}
which may be called the partial inversion of the state \cite{Bloch-N, State-inversion}. According to Corollary \ref{Cor-sym}, if $\rho_{AB}^{(1)}$ is separable, then $\hat{S}_A \otimes \mathds{1} (\rho_{AB}^{(1)})$ must be positive which leads to
\begin{equation}
|t_1|+|t_2|+|t_3| \leq \frac{4}{9} \; . \label{rho1-symm}
\end{equation}
Therefore, $\rho_{AB}^{(1)}$ is entangled when $|t_1|+|t_2|+|t_3| > \frac{4}{9}$ which consistent with the PPT criterion.

Further, our method could make the separable decomposition of $\rho_{AB}^{(1)}$ under the condition of equation (\ref{rho1-symm}). The equation (\ref{rho1-symm}) may be reexpressed as
\begin{equation}
\left(\frac{2}{3}\right)^2 \cdot \left(\frac{2}{3}\right)^2 \geq (|t_1|+ |t_2|+ |t_3|)^2\; .
\end{equation}
According to the positive semidefinite condition of the density matrices in $\mathcal{S}_3^{(A,B)}$, the decomposed local states must be of the following form
\begin{equation}
\rho^{(A,B)} = \frac{1}{3}\mathds{1} + \frac{r}{2}(\sin\theta \cos\phi \lambda_1 + \sin\theta \sin\phi \lambda_2+ \cos\theta \lambda_3) \; , \; r \leq \frac{2}{3} \; . \label{Example-II-local-req}
\end{equation}
The obtained decomposition is similar as that of equations (\ref{decom-24-rp}) and (\ref{decom-24-sp}), i.e.
\begin{equation}
M_{rp} = \frac{1}{2}
\begin{pmatrix}
  \alpha_1 & -\alpha_1 & -\alpha_1 & \alpha_1 \\
  -\alpha_2 & -\alpha_2 & \alpha_2 & \alpha_2 \\
  -\alpha_3 & \alpha_3 & -\alpha_3 & \alpha_3
\end{pmatrix}\; , \; M_{sp} = \frac{1}{2}
\begin{pmatrix}
  \beta_1 & -\beta_1 & -\beta_1 & \beta_1 \\
  -\beta_2 & -\beta_2 & \beta_2 & \beta_2 \\
  -\beta_3 & \beta_3 & - \beta_3 & \beta_3
\end{pmatrix} \; , \label{rho1-decomp}
\end{equation}
where $t_{\mu} = \alpha_{\mu} \beta_{\mu}$, and the norms $|\vec{r}_i|^2 = \sum_{\mu=1}^3 \alpha_{\mu}^2=4/9$ and $|\vec{s}_i|^2 = \sum_{\mu=1}^3 \beta_{\mu}^2 = 4/9$ according to equation (\ref{Example-II-local-req}). Our results show that when (compared to equation (\ref{Example-II-1-other-res}))
\begin{equation}
||\mathcal{T}||_{\mathrm{KF}} = |t_1|+|t_2|+|t_3| \leq \frac{4}{9} \; ,
\end{equation}
$\rho_{AB}^{(1)}$ is separable and the separable decomposition of $\rho_{AB}^{(1)}$ can be written explicitly from $M_{rp}$ and $M_{sp}$ in equation (\ref{rho1-decomp}) with $\vec{r}_i$ and $\vec{s}_i$ being similar to that of equations (\ref{Example-I-detail-1},\ref{Example-I-detail-2}).

For the state $\rho_{AB}^{(2)}$ with $\sqrt{t_1^2 + t_2^2 + t_3^2} \leq 4/9$, the subspaces spanned by left and right singular vectors are different, i.e., $\mathcal{S}_3^{(A)} = \mathrm{span} \{\lambda_1,\lambda_2,\lambda_3\}$, $\mathcal{S}_3^{(B)} = \mathrm{span} \{\lambda_1,\lambda_4,\lambda_6\}$. The positive semidefinite condition for Bloch vectors $\rho^{(B)} = \frac{1}{3}\mathds{1} + \frac{1}{2}( x \lambda_{1} +  y \lambda_4 + z \lambda_6)$ is \cite{Bloch-N}
\begin{align}
x^2 + y^2 + z^2 -3xyz & \leq \frac{4}{9} \; , \label{rho2-region-1} \\
x^2+y^2+z^2 & \leq \frac{4}{3}\; . \label{rho2-region-2}
\end{align}
The convex hull formed by equations (\ref{rho2-region-1}, \ref{rho2-region-2}) has the shape of fully filled rice dumpling (a traditional Chinese food) with four vertices coincident to a regular tetrahedron
\begin{equation}
(\frac{2}{3},\frac{2}{3},\frac{2}{3})\; , \; (-\frac{2}{3},-\frac{2}{3},\frac{2}{3})\;, \; (-\frac{2}{3},\frac{2}{3},-\frac{2}{3})\; ,\; (\frac{2}{3},-\frac{2}{3},-\frac{2}{3})\;.
\end{equation}
From Corollary \ref{Cor-uncertainty} and equation (\ref{rho2-region-2}), we have
\begin{equation}
\left(\sum_{\mu=1}^3 \mathcal{E}_{\mu} (A) \right)\cdot \frac{4}{3} \geq ||\mathcal{T}||^2_{\mathrm{KF}} = (|t_1|+|t_2| + |t_3|)^2\; . \label{Example-II-2-8}
\end{equation}
From $\sqrt{t_1^2 + t_2^2 + t_3^2} \leq 4/9$, we can get $(|t_1|+|t_2| + |t_3|)^2 \leq \frac{16}{27}$. Taking this into equation (\ref{Example-II-2-8}), we get $\sum_{\mu=1}^3 \mathcal{E}_{\mu} (A) \leq \frac{4}{9}$. Our result predicts that the state $\rho_{AB}^{(2)}$ admits the following separable decomposition
\begin{align}
M_{rp} & = M_r \cdot D^{\frac{1}{2}}_{p} =
\begin{pmatrix}
\alpha_1 & \alpha_1 & -\alpha_1 & -\alpha_1 \\
\alpha_2 & -\alpha_2 & -\alpha_2 & \alpha_2 \\
\alpha_3 & -\alpha_3 & \alpha_3 & -\alpha_3
\end{pmatrix} \cdot
\begin{pmatrix}
\frac{1}{2} & 0 & 0 & 0 \\
0 & \frac{1}{2} & 0 & 0 \\
0 & 0 & \frac{1}{2} & 0 \\
0 & 0 & 0 & \frac{1}{2}
\end{pmatrix}\; , \\
M_{sp} & = M_s\cdot D^{\frac{1}{2}}_{p} =
\begin{pmatrix}
\beta_1 & \beta_1 & -\beta_1 & -\beta_1 \\
\beta_2 & -\beta_2 & -\beta_2 & \beta_2 \\
\beta_3 & -\beta_3 & \beta_3 & -\beta_3
\end{pmatrix} \cdot
\begin{pmatrix}
\frac{1}{2} & 0 & 0 & 0 \\
0 & \frac{1}{2} & 0 & 0 \\
0 & 0 & \frac{1}{2} & 0 \\
0 & 0 & 0 & \frac{1}{2}
\end{pmatrix} \; ,
\end{align}
where $\alpha_{\mu} \beta_{\mu}=t_{\mu}$. Therefore, $\rho_{AB}^{(2)}$ is separable when $\beta_{\mu} = t_{\mu}/\alpha_{\mu}$  satisfy equations (\ref{rho2-region-1}, \ref{rho2-region-2})
\begin{align}
\frac{t_1^2}{\alpha_1^2} + \frac{t_2^2}{\alpha_2^2} + \frac{t_3^2}{\alpha_3^2} - 3\frac{t_1t_2t_3}{\alpha_1\alpha_2\alpha_3} \leq \frac{4}{9} \; , \; \mathrm{and} \; \frac{t_1^2}{\alpha_1^2} + \frac{t_2^2}{\alpha_2^2} + \frac{t_3^2}{\alpha_3^2} \leq \frac{4}{3} \; . \label{rho2-sep-reg}
\end{align}
Here the Bloch vectors' norm of $\rho^{(A)}$ should be $\alpha_1^2 + \alpha_2^2 + \alpha_3^2 \leq \frac{4}{9}$. Equation (\ref{rho2-sep-reg}) covers the region $\sqrt{t_1^2 + t_2^2 + t_3^2} \leq 4/9$, so $\rho_{AB}^{(2)}$ is separable for all values of $t_i$ in equation (\ref{Example-II-2-p-c}).

In the above examples, we use Corollaries \ref{Cor-uncertainty} and \ref{Cor-sym} to detect the entanglement of mixed states and compare the results with that of PPT and Bloch representation criteria. (Many other criteria reduce to these two criteria for normal form states \cite{Bloch-Rep}.) Then, we explicitly construct the separable decompositions for the bipartite mixed states based on the separability criteria of Corollaries \ref{Cor-uncertainty} and \ref{Cor-sym}. Here the decomposition is done within the left and right singular vector spaces (i.e., $\mathcal{S}_l^{(A)}$ and $\mathcal{S}_l^{(B)}$), where, generally, dimensions out of $\mathcal{S}_l^{(A)}$ or $\mathcal{S}_l^{(B)}$ may also be admissible. Full analysis  of the symmetries and volumes of the general convex hulls of the local Bloch vectors would give the necessary and sufficient condition for separability.

\section{Discussion}

We have shown that, by factoring the correlation matrix into two matrices, practical entanglement criteria are derived within the scheme developed in \cite{Separability-QL}. With these criteria, one can analyze the magnitudes and symmetries of the Bloch vectors of the decomposed local states, which is unreachable for other criteria based on matrix norms. Furthermore, our method provides a practical way to construct the separable decomposition of mixed bipartite states analytically, rather than merely realizing the decompositions numerically. Since the Bloch vectors of local states are distributed over a high dimensional convex hull closely related to the steering ellipsoid in quantum steering \cite{Steering-ellip}, our method might play an essential role in the studies of variant nonlocal phenomena, e.g., mixed state entanglement, quantum steering, and Bell nonlocality.

\section*{Acknowledgements}

\noindent This work was supported in part by the Ministry of Science and Technology of the Peoples' Republic of China(2015CB856703); by the Strategic Priority Research Program of the Chinese Academy of Sciences(XDB23030100); and by the National Natural Science Foundation of China(NSFC) under the grants 11375200 and 11635009.

\end{document}